\documentclass[letterpaper,12pt]{article}   
\usepackage{osajnl2} 
\usepackage{psfrag}
\usepackage{color}
\usepackage[ansinew]{inputenc}
\usepackage{graphicx}

\definecolor{gris}{gray}{0.3}
\definecolor{vertfonce}{rgb}{0,0.5,0}
\definecolor{vert}{cmyk}{0.7,0,.7,0.5}
\definecolor{rouge}{cmyk}{0,1,1,.3}
\definecolor{bleu}{cmyk}{1,0,0,.3}
\definecolor{kaki}{cmyk}{0,0,1,.3}

\begin{document}
%
%
\title{Low power inelastic light scattering at small detunings in silicon wire waveguides at telecom wavelengths}
%
%
%
\author{St\'ephane Clemmen $^{1,*}$, Antony Perret $^2$,  Jassem Safioui $^{3}$, Wim Bogaerts $^4$, Roel Baets $^4$, Simon-Pierre Gorza $^3$,  Philippe Emplit $^3$ and Serge Massar $^1$ }

\address{$^{1,5}$ Laboratoire d'Information Quantique, CP 225,  Universit\'{e} Libre de Bruxelles (U.L.B.),
 Boulevard du Triomphe, B-1050 Bruxelles, Belgium}
\address{$^2$ D\'epartement de Physique, \'{E}cole normale sup\'erieure, 24 rue Lhomond, 75005 Paris, France}
\address{$^3$ Service OPERA-Photonique, CP 194/5, Universit\'e Libre de Bruxelles (U.L.B.), avenue F.D. Roosevelt 50,  1050 Brussels, Belgium}
\address{$^4$ Photonics Research Group, INTEC-department, Ghent University - IMEC,  Sint-Pietersnieuwstraat 41, 9000 Gent, Belgium}
\address{$^5$ Currently with School of Applied and Engineering Physics, Cornell University, Ithaca, New York 14853, USA}
\address{$^*$Corresponding author : sclemmen@ulb.ac.be}
%
%
%
%
\begin{abstract}

When a pump beam is propagating through a silicon nanophotonic waveguide, a very small fraction of the light is scattered to other frequencies. At very low intensity, the amount of scattered light is proportional to the power of the pump beam.  We show that the scattering intensity increases linearly within the temperature range 300--575~K and that the photon flux decreases as the inverse of the frequency detuning  $\nu$ over the investigated bandwidth $0.4 \: \textrm{THz}<|\nu|<2.5 \: \textrm{THz}$. 
The simplest interpretation of these observations is that the pump beam is scattered on a 1 dimensional thermal bath of excitations.
Finally, the implications of this scattering process for quantum optics applications of silicon nanophotonic structures are discussed.
\hspace{-0.3cm} 
\end{abstract}
%
%
%
%
%
\section{Introduction}

Light propagation in silicon wire waveguides (SWW) at telecommunication wavelengths is highly complex. Indeed such waveguides exhibit an intrinsic Kerr non linearity, a Raman scattering with a narrow peak at $15.6$~THz, linear absorption, as well as two photon absorption which generate free carriers, in turn responsible for free carriers absorption. Reviews presenting these processes, as well as some of their consequences can be found in~\cite{lin2007nonlinear,claps2002observation,turnerfoster2010ultrashort,solli2009inverse}. 

Recently, SWW have been investigated as a promising source of correlated photon pairs~\cite{sharping2006generation, lin2006silicon, takesue2007entanglement, takesue2008generation, harada2008generation, clemmen2009continuous, harada2010}. These are generated via a four wave mixing process in which the third order non linearity of the medium converts two photons from the pump beam into two correlated photons at frequency detunings $\pm \nu$ from the pump frequency. However, experimental studies of photon pairs generation in SWW have revealed an unexpected photon noise source~\cite{sharping2006generation, clemmen2009continuous}. In particular, the experiment reported in Ref.~\cite{clemmen2009continuous} was done in CW regime at very low power and was much more sensitive to noise than earlier works. A large reduction from the predicted signal to noise ratio was then measured (11.3 instead of 69). Some potential sources of noise in SWW were theoretically studied in~\cite{lin2006silicon} but these do not seem to explain the observations of~\cite{clemmen2009continuous}.  We note that a similar unexpected source of noise has also been observed in hydrogenated amorphous silicon waveguides~\cite{Clemmen:10}.

Here we report on a detailed experimental study of the inelastic light scattering that occurs in SWW for frequency detunings well below the Raman peak, and we demonstrate that this scattering process is responsible for the low signal to noise ratio reported in~\cite{clemmen2009continuous}. The spectrum of the scattered light and its dependence on the waveguide temperature 
is most simply interpreted as scattering on a 1 dimensional thermal bath of excitations, involving both absorption  and  emission of excitations.
Finally, time resolved measurements clearly demonstrate that this scattering is not related to the free carriers density.

\section{Experimental setup}

\begin{figure}
\begin{center}
\psfrag{laser}[cc][cc][1][0]{\footnotesize{laser}}
\psfrag{edfa}[cc][cc][1][0]{\footnotesize{edfa}}
\psfrag{p-c}[cc][cc][1][0]{\footnotesize{pc}}
\psfrag{dmux}[cc][cc][1][0]{\footnotesize{dmux}}
\psfrag{power-m}[cc][cc][1][0]{\footnotesize{P-meter}}
\psfrag{sss}[cc][cc][1][0]{\footnotesize{ssw}}
\psfrag{f-f}[cc][cc][1][0]{\footnotesize{flip-flop}}
\psfrag{col}[cc][cc][1][0]{\footnotesize{col}}
\psfrag{apd}[cc][cc][1][0]{\footnotesize{apd}}
\psfrag{TDC}[cc][cc][1][0]{\footnotesize{tdc}}
\psfrag{K}[cc][cc][1][0]{\scriptsize{$\sharp$}}
\psfrag{d-e2}[cc][cc][1][0]{\scriptsize{time}}
\psfrag{hist}[cc][cc][1][0]{}
\psfrag{d-e}[cc][cc][1][0]{\footnotesize{delay}}
\psfrag{bpf}[cc][cc][1][0]{\footnotesize{bpf}}
\psfrag{bbf}[cc][cc][1][0]{\footnotesize{bbf}}
\psfrag{t-f}[cc][cc][1][0]{\footnotesize{tf}}
\psfrag{p-g}[cc][cc][1][0]{\footnotesize{pg}}
\psfrag{basicsetup}[cc][cc][1][0]{\footnotesize{pair}}
\psfrag{bas2}[cc][cc][1][0]{\footnotesize{generation}}
\psfrag{analysis}[cc][cc][1][0]{\footnotesize{analysis}}
\psfrag{ana2}[cc][cc][1][0]{\footnotesize{setup}}
\includegraphics[width=8cm]{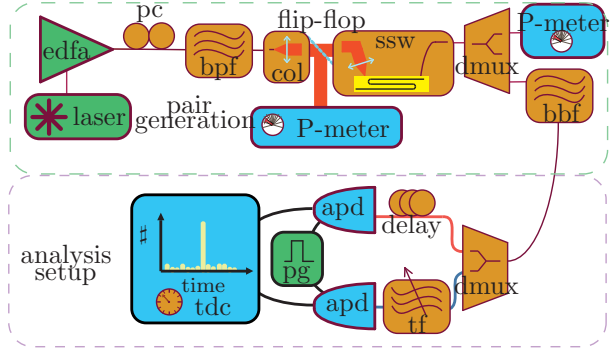}
\caption{Experimental setup. Photon flux is generated in the top part of the setup and the bottom part is used to analyse this flux in the Stokes and Anti-Stokes bands. Laser: CW laser source at 1539.8~nm, EDFA: erbium-doped fiber amplifier, pc: polarization controller, bpf: band pass filter to clean the pump beam, col: collimation lens, flip flop: mirror used to measure the input power, SWW: silicon wire waveguide, demux: demultiplexer, tf: tunable filter, apd: avalanche photodiode, pg, pulse generator, tdc: time to digital converter + computer.   
\label{fig:setup1}}
\end{center}
\end{figure}

Our experiments were performed with silicon wire waveguides (cross section $500\times220$\,nm) with a lower cladding of silica and air as an upper cladding. These waveguides are 11.2~mm long and are characterized by a group velocity dispersion $\beta_2 =-1.5\pm0.5$~ps$^{2}$ m$^{-1}$ at a wavelength of $1.5 \: \mu$m and a third order nonlinear coefficient $\gamma \approx 300 \textrm{ W}^{-1} \textrm{m}^{-1}$ (assuming an effective area of $0.064\: \mu\textrm{m}^{2}$ and a Kerr coefficient of~\cite{tsang2008nonlinear} $4.7 \times 10^{-18} \: \textrm{m}^2/\textrm{W} $).

The experimental setup is depicted in Fig~\ref{fig:setup1} and can be divided into two main parts. The first part (top pannel in Fig~\ref{fig:setup1}) is used to generate the scattered flux of photons in the waveguide. The second part (bottom pannel) is an analysing system to measure the photon flux power, the spectrally resolved photon flux, and the spectrally resolved coincidences. The pump beam is a CW laser beam at 1539.8~nm (Agilent 81600b amplified by a Keopsys ErYb-doped fiber amplifier) which has been cleaned on Stokes and anti-Stokes frequencies using a bandpass filter (BPF). This BPF consists of fiber Bragg gratings, circulators and 100~GHz commercial add \& drop filters. It has an extinction larger than $150$~dB outside of the pump band ($1539.8 \pm 0.8$~nm). Note that to reduce as much as possible Raman scattering in the fiber, the pigtail of the last component of the BPF is only 10~cm long. After the BPF, the light is first collimated, then injected into the SWW by focussing the beam onto a grating coupler. At the output of the waveguide, the light is coupled into a standard optical fiber by means of a second grating coupler. The input and output coupling losses as well as the linear propagation loss in the waveguide have been measured to be $6\pm0.5$~dB and $2.5\pm0.5$~dB, respectively. After outcoupling, the pump beam is separated by a demultiplexer and sent to a powermeter. In the second output port of the demultiplexer, the remaining part of the pump beam is removed by a bandblock filter (bbf) made up of fiber Bragg gratings (extinction greater than 150~dB in the pump band). The Stokes (1542-1558\,nm) and anti-Stokes bands (1522-1538\,nm) are then separated by a second demultiplexer. Measurements of the photon fluxes in these two bands are carried out with single photon detectors. These detectors are avalanche photodiodes (APD - ID201 from ID quantique) operating in Geiger mode at 100~kHz with a gate duration of 100~ns synchronized with a pulse delay generator (Standford DG535). The efficiency of the detectors are $10\pm 1\%$ (Stokes) and $15\pm1 \%$ (anti-Stokes) while dark counts are $ 805 \pm 3$~Hz and $155 \pm 1$~Hz, respectively. The temporal resolution of the detectors is limited by the time-to-dital converter (tdc) and is less than 1~ns. For the spectrally resolved measurements, a tunable filter (gaussian transmission with a full width at half maximum of 1.5~nm - Newport TBF-1550-1.0) is added in the Stokes arm after the last demultiplexer.

The temperature of the SWW can be varied from 300~K (room temperature) to 575~K thanks to a cartridge heater embedded in the aluminium block holding the waveguide and its temperature is monitored with a thermocouple. Except when mentioned explicitly, all measurements are carried out at room temperature ($\approx$300~K).

\section{Experimental results}

\begin{figure}\begin{center}
\psfrag{pumppower}[cc][cc][1][0]{\footnotesize{Pump power {in waveguide} (mW)}}
\psfrag{flux}[cc][cc][1][0]{\footnotesize{Generated photon flux ($\times 10^8$ Hz)}}
\psfrag{x0}[cc][cc][1][0]{\footnotesize{0}}
\psfrag{x1}[cc][cc][1][0]{\footnotesize{0.5}}
\psfrag{x2}[cc][cc][1][0]{\footnotesize{1}}
\psfrag{x3}[cc][cc][1][0]{\footnotesize{1.5}}
\psfrag{x4}[cc][cc][1][0]{\footnotesize{2}}
\psfrag{x5}[cc][cc][1][0]{\footnotesize{2.5}}
\psfrag{x6}[cc][cc][1][0]{\footnotesize{3}}
\psfrag{Y1}[cc][cc][1][0]{\footnotesize{0}}
\psfrag{Y2}[cc][cc][1][0]{\footnotesize{1}}
\psfrag{Y3}[cc][cc][1][0]{\footnotesize{2}}
\includegraphics[width=7.5cm]{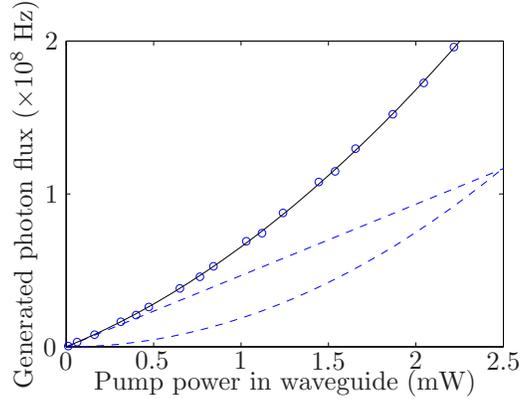}
\caption{Experimentally recorded photons flux generated in the Stokes band (1541.5-1558.5~nm) as a function of the pump power. Solid line: second order polynomial fit $\Phi=aP^2+bP$, dotted curves: linear ($bP$) and quadratic ($aP^2$) contributions.
}
\label{fig:flux}
\end{center}\end{figure}

The power of the scattered light in the Stokes band was first measured as a function of the input power of the pump laser. As can be seen in Fig.~\ref{fig:flux}, the power in the Stokes band increases with the pump power ($P$) and exhibits a linear as well as a quadratic dependence with $P$. The quadratic contribution is well known~\cite{clemmen2009continuous} and can be attributed to the nonlinear third order Kerr effect. Indeed, when a continuous wave is propagating in a medium, the  instantaneous third order nonlinear effect is responsible for the generation of photon pairs through degenerated four-wave mixing process. The flux of photon pairs emitted over a small frequency interval $\Delta \nu$ at detuning $\nu$ from the pump frequency is given by~\cite{brainis2009four}
\begin{equation}
\Phi =
 \int_{\Delta \nu}   \left| \gamma P L \:  \textrm{sinc} \left[ \beta_2^{\frac{1}{2}} 2 \pi \nu  L  \left( \frac{\beta_2 (2 \pi \nu)^2}{4}  +\gamma P \right)^{\frac{1}{2}} \right]   \right| ^2  \:  \textrm{d} \omega    
\label{eq:flux}
\end{equation}
and thus increases quadratically with $P$, while its spectrum $\Phi(\nu)$ has a sinc shape for small frequency interval $\Delta\nu$.

Beside the quadratic contribution to the photon flux, a linear contribution is clearly visible in Fig.~\ref{fig:flux}. This later contribution cannot be explained by nonlinear losses since the power in the waveguide is too small for these losses to be significant. This has been experimentally verified by measuring the propagation losses in the SWW for each data points reported in Fig.~\ref{fig:flux}. The highest relative variation of the ratio of the output to the input power was $2.5\%$ and is attributed to small coupling loss variations. 
 
Raman scattering occurring in the filtering line between the SSW and the detectors is a potential source of photons since this filter is made up of standard optical fiber components. 
The measurement of this contribution to the total flux was performed by replacing the SWW by a fiber-attenuator which introduces the same amount of power attenuation as the SWW. At an input pump power of 1.25~mW (half the maximum power considered in our experiments), the generated photon flux was found to be less than $12\%$ of the total flux in the spectral bandwidth previously considered. Raman scattering in the filtering line cannot therefore account for the difference between the experimentally recorded photon flux and the quadratic law given by Eq.\ref{eq:flux}.

\begin{figure*}  
%
\hspace{1cm}
\psfrag{Pairflux}[cb][ct][1][0]{\footnotesize{Pair flux (kHz)}}
\psfrag{Photonflux}[cb][ct][1][0]{\footnotesize{Photon flux (MHz)}}
\psfrag{wavdetuning}[ct][cb][1][0]{\scriptsize{Wavelength detuning (nm)}}
\psfrag{w0}[cc][cc][1][0]{\footnotesize{0}}
\psfrag{w1}[cc][cc][1][0]{\footnotesize{}}
\psfrag{w2}[cc][cc][1][0]{\footnotesize{10}}
\psfrag{w3}[cc][cc][1][0]{\footnotesize{}}
\psfrag{w4}[cc][cc][1][0]{\footnotesize{20}}
\psfrag{x0}[cb][cc][1][0]{\footnotesize{0}}
\psfrag{x2}[cc][cc][1][0]{\footnotesize{2}}
\psfrag{x4}[cc][cc][1][0]{\footnotesize{4}}
\psfrag{x6}[cc][cc][1][0]{\footnotesize{6}}
\psfrag{ k0}[cb][cc][1][0]{\textcolor{vertfonce}{\footnotesize{0}}}
\psfrag{ k2}[cc][cc][1][0]{\textcolor{vertfonce}{\footnotesize{20}}}
\psfrag{ k4}[cc][cc][1][0]{\textcolor{vertfonce}{\footnotesize{40}}}
\psfrag{ k6}[cc][cc][1][0]{\textcolor{vertfonce}{\footnotesize{60}}}
\psfrag{y0}[cb][cc][1][0]{\footnotesize{0}}
\psfrag{y2}[cc][cc][1][0]{\footnotesize{.1}}
\psfrag{y4}[cc][cc][1][0]{\footnotesize{.2}}
\psfrag{y6}[cc][cc][1][0]{\footnotesize{.3}}
\psfrag{ p0}[cb][cc][1][0]{\textcolor{vertfonce}{\footnotesize{0}}}
\psfrag{ p2}[cc][cc][1][0]{\textcolor{vertfonce}{\footnotesize{.5}}}
\psfrag{ p4}[cc][cc][1][0]{\textcolor{vertfonce}{\footnotesize{1}}}
\psfrag{ p6}[cc][cc][1][0]{\textcolor{vertfonce}{\footnotesize{1.5}}}
\psfrag{z0}[cb][cc][1][0]{\footnotesize{0}}
\psfrag{z2}[cc][lc][1][0]{\footnotesize{20}}
\psfrag{z4}[cc][lc][1][0]{\footnotesize{40}}
\psfrag{z6}[cc][lc][1][0]{\footnotesize{60}}
\psfrag{ r0}[cb][cc][1][0]{\textcolor{vertfonce}{\footnotesize{0}}}
\psfrag{ r2}[cc][rc][1][0]{\textcolor{vertfonce}{\footnotesize{1}}}
\psfrag{ r4}[cc][rc][1][0]{\textcolor{vertfonce}{\footnotesize{2}}}
\psfrag{ r6}[cc][rc][1][0]{\textcolor{vertfonce}{\footnotesize{3}}}
\psfrag{aa}[cb][cc][1][0]{\footnotesize{}}
\psfrag{bb}[cc][cc][1][0]{\footnotesize{}}
\psfrag{cc}[cc][cc][1][0]{\footnotesize{}}
\includegraphics[width=14cm]{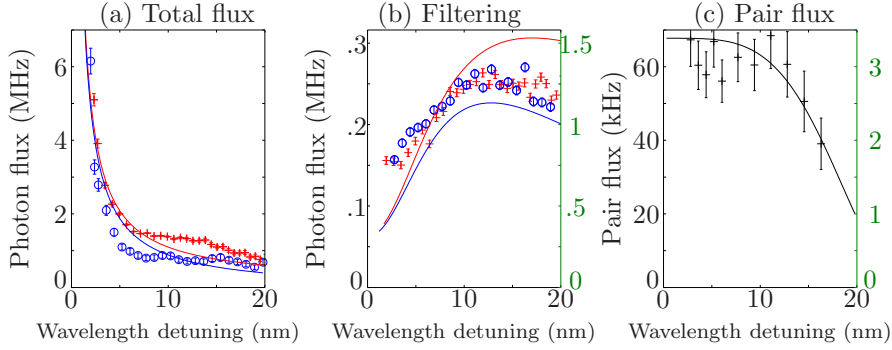}
\setlength{\unitlength}{1cm}
\put(-11.6,4){\footnotesize{(a) Total flux}}
\put(-7.9,4){\footnotesize{(b) Filtering}}
\put(-3.7,4){\footnotesize{(c) Pair flux}}
\caption{{a)~Generated photon flux in the Stokes ({\color{red} +}) and anti-Stokes ({\color{blue} o}) bands (1541.5-1558.5~nm and 1522-1528~nm, respectively), for $P=250~\mu$W. The solid curves are theoretical fits with a Bose-Einstein distribution  (see Eq.\,\ref{eq:Pscat}). b)~Photon flux generated in the filtering line itself at Stokes ({\color{red} +}) and anti-Stokes ({\color{blue} o}) frequencies for input power $P=1.25$~mW. Solid curves: fit following the Raman noise in a silica fiber.  
c)~Photon pair flux generated in the silicon waveguide (+) and fit following eq.~\ref{eq:flux} (curve), input power: 1.25\,mW. Error bars are calculated from statistical error as well as error on out-coupling losses. In these figures, the flux for each data point is corrected by subtracting the loss spectrum and dark counts from the detectors. The experiments have been performed at different input powers to ensure similar statistical errors. The right axis (coloured in green) refers to the actual power while the left axis rescales the data to input power $P=250$\,$\mu$W to enable comparison with the spectrum reported in Fig.(a) (see discussion in main text).} }
\label{fig:spectrum}
\end{figure*}

To clearly identify the physical process responsible for the linear contribution reported in Fig.~\ref{fig:flux}, we experimentally studied the spectral characteristics of the generated photon fluxes in the Stokes and anti-Stokes bands. The spectrally resolved photon flux in these two bands, recorded at a pump power of 250~$\mu$W, are shown in Fig.\ref{fig:spectrum} (a). As can be seen, the photon fluxes rapidly decrease with the frequency detuning $|\nu|$. These spectra are in agreement with a Bose-Einstein law (solid lines in Fig.\ref{fig:spectrum} (a)
which are given by eq. (\ref{eq:Pscat})), and support the assumption that photons are scattered by thermal excitations. Moreover, the flux in the Stokes band is larger than its anti-Stokes counterpart. This is expected if the Stokes band is associated with emission of excitations and the anti-Stokes band to absorption of excitations. The spectrum of Fig.\ref{fig:spectrum} (a) was then compared with the spectrally resolved photon flux generated in the fiber, i.e. when the waveguide is replaced by the fiber  attenuator, see Fig. \ref{fig:spectrum}(b). The later data were recorded at a power of 1.25~mW to ensure similar statistical errors. 
The origin of the scattered photons in the fiber is expected to be Raman scattering, and is therefore expected to scale linearly with input power.
Upon rescaling by a factor 
$250 \mu{\textrm{W}}/1.25 {\textrm{mW}}=0.2$ to compare with panel (a) (left axis in Fig.\ref{fig:spectrum}(b)) it appears that within the studied spectral band, photons generated by Raman scattering in the filtering line account for maximum $25\%$ of the spectral flux up to a detuning of 20~nm. The measurements in Fig.\ref{fig:spectrum} (a) where further compared with the spectrum of the correlated photon pairs generated in the SWW. This spectrum was obtained  from a coincidence measurement at a pump power of 1.75~mW and is shown in Fig.\ref{fig:spectrum}(c). It is in agreement with the sinc variation predicted from Eq.\ref{eq:flux}. This spectrum is expected to scale quadratically with input power.
Upon rescaling by a factor 
$(250 \mu{\textrm{W}}/1.75 {\textrm{mW}})^2 \simeq 0.02$ to compare with panel (a) (left axis in Fig.\ref{fig:spectrum}(c))
it appears that the flux coming from correlated photons pairs is negligible in comparison with the total flux.
In conclusion, the linear contribution of the power dependence of the photon flux generated in the SWW can therefore be attributed to a scattering process whose spectrum is well represented by the measurements reported in Fig.\ref{fig:flux}(a).

\begin{figure}  
\begin{center}
\psfrag{labelx}[cc][cu][1][0]{\footnotesize{Temperature (K)}}
\psfrag{labely}[cc][cc][1][0]{\footnotesize{Generated flux (MHz)}}
\psfrag{X0}[cc][cc][1][0]{\footnotesize{300}}
\psfrag{X1}[cc][cc][1][0]{\footnotesize{400}}
\psfrag{X2}[cc][cc][1][0]{\footnotesize{500}}
\psfrag{X3}[cc][cc][1][0]{\footnotesize{600}}
\psfrag{Y1}[cc][cc][1][0]{\footnotesize{50}}
\psfrag{Y2}[cc][cc][1][0]{\footnotesize{100}}
\psfrag{Y3}[cc][cc][1][0]{\footnotesize{150}}
\includegraphics[width=7.5cm]{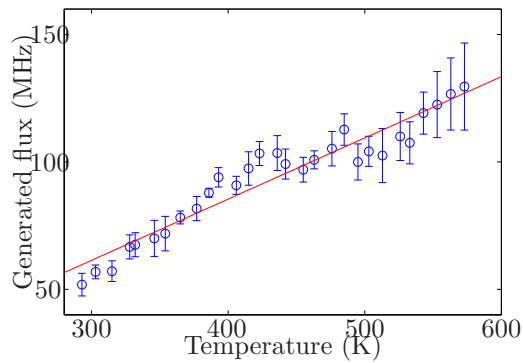}
\caption{Flux generated in the anti-Stokes band as a function of the SWW temperature \textbf{ ({\color{blue} o}) } and linear fit 
\textbf{ ({\color{red} -}) }.}  
\label{fig:thermal}
\end{center}\end{figure}

In order to confirm that the scattering process involves interaction with thermal excitations in the SWW, we have measured the evolution of the flux in the anti-Stokes band as a function of the temperature of the waveguide. In this experiment, the input pump power was set to a value of 500~$\mu$W and the total flux generated in the spectral band from 0.4 to 2.5~THz from the pump frequency was recorded. The temperature dependence of the photon flux is shown in Fig.\ref{fig:thermal} and is found to be linear in the temperature range extending from 300~K to 575~K. Note that this flux is corrected for losses and detector efficiency.

\begin{figure}
\begin{center} 
\psfrag{ylabel}[cc][cc][1][0]{\footnotesize{Normalized flux (linear scale)}}
\psfrag{xlabel}[cc][cc][1][0]{\footnotesize{Time [ns]}}
\psfrag{x0}[cc][cc][1][0]{\footnotesize{-1}}
\psfrag{x1}[cc][cc][1][0]{\footnotesize{0}}
\psfrag{x2}[cc][cc][1][0]{\footnotesize{1}}
\psfrag{x3}[cc][cc][1][0]{\footnotesize{49}}
\psfrag{x4}[cc][cc][1][0]{\footnotesize{50}}
\psfrag{x5}[cc][cc][1][0]{\footnotesize{51}} 
\psfrag{Y0}[cc][cc][1][0]{\footnotesize{0}} 
\psfrag{pulseduration}[cc][cc][1][0]{\footnotesize{50 ns}}
\includegraphics[width=7.5cm]{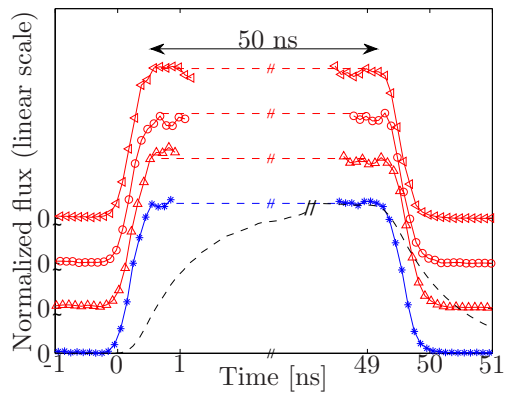}
\caption{Emitted flux response time. 
{Measured temporal profile of the input pump pulse ({\color{blue} *})} and corresponding total photon flux emitted at both Stokes and anti-Stokes frequencies (red - 3 top curves) for 300~$\mu$W~({\color{red} $\Delta$}), 1.25~mW~({\color{red} $\circ$}), and 2.5~mW~({\color{red} $\triangleleft$}) peak pump power. %
The black dotted curve shows the temporal dynamic of the carrier density due to the laser pulse assuming a carrier lifetime of 1~ns.
Rise/fall time ($10\%$-$90\%$) is the same (450~ps) for both the laser and the 3 scattered fluxes. The corresponding rise/fall time ($10\%$-$90\%$)  of  the carrier population is 2.3~ns assuming a 1~ns lifetime. The scattered flux is thus independent of the carrier population.
}
\label{fig:timeresponse}
\end{center}\end{figure}

%
Free carrier generation is known to play a major role in the optical properties of silicon wire waveguides at a wavelength of 1.5~$\mu$m and could potentially be involved in the inelastic photon scattering reported in this work. Free carries arise when electron-hole pairs are created through linear and nonlinear absorption of photons. The free carrier density is therefore time dependent when pulses are launched in SWW and its effects on light scattering can thus be investigated by time resolved measurements. To study the possible role of free carriers, the CW input beam was cut into square pulses. To achieve this, the output of a pulse pattern generator(Agilent 81110A) was sent through a non linear diode that induces sharp rising slopes of the electrical pulses. These electrical pulses drive an integrated Lithium-Niobate intensity modulator, thereby temporally shaping the input beam into 50~ns square pulses at a 2~MHz repetition rate.
 The pulse duration and the repetition rate were chosen such that the carrier population can fully build up during each pulse and completely decay between two pulses (the carrier lifetime is $\approx1$\,ns). Fast superconducting single photon detector (Scontel detector - jitter around 40~ps) and high performance time-to-digital converter (Agilent Acqirisis system - 5~ps resolution) ensure a time resolution of 80~ps. This overall temporal resolution is limited by the pulse pattern generator, but is largely sufficient to reveal any free carriers effects. 
The peak pump power has been adjusted so that the linear part of the photon flux is the dominant source of photon generation away from the pump frequency ($P=0.3,\  1.25,\ 2.5$~mW). This power level also keeps the average number of scattered photons per pulse much lower than 1 to avoid missing late events because of the dead time of the detector. The laser pulses have been measured by the same setup as for the scattered photon flux and are shown in Fig.\,\ref{fig:timeresponse} (bottom curve) together with the photon flux for different input peak powers. As can be seen, the temporally resolved flux of the scattered light is independent of the input power $P$ and exhibits a rise time smaller or equal than 100~ps.      
This time scale is much faster than the typical values associated with the carriers dynamics in similar waveguides. This latter time ranges between 1~ns and 4~ns, 1~ns being the commonly accepted value and 4~ns, the carrier lifetime measured in the waveguide used for our experiments at the same power level (note that this value is compatible with~\cite{turnerfoster2010ultrashort}). These results clearly show that the inelastic scattering associated with the linear contribution reported in Fig.\ref{fig:flux} is not related to the free carriers density. 

\section{Discussion}

We have presented evidence for broadband low power light scattering in SWW that exhibits a strong thermal and spectral dependence. In our opinion, the simplest explanation for these observations is inelastic scattering of the pump beam on a thermal distribution of excitations. This model predicts that the probability for a pump photon at frequency $\nu_0$ to be scattered to frequency $\nu_0+\nu$ is proportional to the number of excitations with frequency $\nu$. 
More precisely, assuming a pump beam of power $P$ propagating in a waveguide of length $L$ at a temperature $T$, 
the power of the light that is scattered in the frequency interval $\Delta \nu$ around $\nu_0+\nu$ should have the form
\begin{equation}
P_{scat}=\kappa \left[\frac{1}{ \exp(\textrm{h} |\nu|/\textrm{k}_{\textrm{b}} T)-1} + \frac{1}{2} \left( 1 - \textrm{sign} (\nu)  \right) \right]   L \: \Delta\nu \: P
\label{eq:Pscat}
\end{equation}
where $\textrm{sign} (\nu)=\pm 1$ corresponds to Stokes and anti-Stokes scattering respectively. In general $\kappa$ can depend on $\nu$, both through the spectral density of excitations and through the spectral dependency of the scattering amplitude. However, the spectrum reported in Fig.\ref{fig:spectrum} (a) is well fitted by eq. (\ref{eq:Pscat}) with $\kappa$ independent of $\nu$ (see solid curves in the figure). This suggests that the scattering is on a 1 dimensional gas of excitations, and that the scattering amplitude varies little with detuning.

We note that within the investigated range of frequency detunings and temperatures the Bose-Einstein distribution 
$\left( \exp(  \textrm{h} |\nu|/ \textrm{k}_{\textrm{b}} T)-1\right)^{-1} $ is well approximated by $\textrm{k}_{\textrm{b}} T / \textrm{h} |\nu|$, leading to a linear temperature dependence and a spectral variation that scales as the inverse of the frequency detuning. Confirmation of the complete Bose-Einstein distribution predicted by eq. ({eq:Pscat}) would require measurements at larger detunings or lower temperature, which is unfortunately not accessible with our setup.

From the data presented in Fig \ref{fig:flux} and Fig.\ref{fig:spectrum} (c), we estimate that for the waveguide studied, the proportionality constant is $\kappa=3.5 \times 10^{-10}   \textrm{cm}^{-1}\textrm{THz}^{-1} \: \pm 30 \% $.

The physical origin of the 1 dimensional thermal gas of excitations on which the photons scatter is unclear. Phonons are a natural candidate. However, the phonon dispersion relation in bulk crystalline silicon (see e.g.~\cite{PhysRevB.50.2221}) does not allow for the scattering observed here. A possible explanation is a modification of the continuous dispersion relation of phonons in bulk silicon either through geometrical effects due to the finite dimension of the waveguide, or because of the presence of defects in the silicon. Explanations based on Raman scattering in the silica substrate beneath the waveguide are probably ruled out because the scattered spectrum would be very different from the observed spectrum. Explanations based on free carriers are also ruled out by our temporal measurements. The fact that a similar scattering was also observed in hydrogenated amorphous-silicon waveguides~\cite{Clemmen:10} constrains possible interpretations of our observations.

The linear broadband scattering studied in the present work is very weak. Nevertheless, it has implications for the future use of SWW in quantum optics. Indeed our work shows that photon pair sources based on SWW can be improved, either by cooling the SWW, or (easier) by  selecting the Stokes and anti-Stokes spectral bands not too close to the pump wavelength.  

We emphasize that notwithstanding the scattering studied in the present paper noise, SWW remain a promising source of photon pairs based on four wave mixing.  Indeed, we have estimated that the signal to noise ratio (SNR) obtained from coincidence measurements on SWW and a comparable silica fiber source (silica fibers are the most studied photon pair sources based on four wave mixing~\cite{fiorentino2002allfiber, takesue2004generation, li2004allfiber}) differ by at least a factor 20. In the case of silica fibers, we made a theoretical estimate based on the  theory presented in~\cite{brainis2007spontaneous}. For the comparison we tried to keep all other factors constant, and in particular assumed zero dispersion for the fiber so that the spectrum of the pair flux is flat, and integrated flux over the same Stokes and anti-Stokes bands as used here. This explains why Raman scattering has plagued photon pair generation experiments based on silica fibers from the onset, but has only been noticed recently in the case of SWW.
\section*{Acknowledgments}
We acknowledge the support of the \textit{Fonds pour la formation \`a la Recherche dans l'Industrie et dans l'Agriculture} (FRIA, Belgium), of the \textit{Interuniversity Attraction Poles Photonics@be Programme} (Belgian Science Policy) under grant IAP6-10,
of the \textit{Fonds pour la Recherche Fondamentale Collective} (FRFC) of the FRS-FNRS under grant number 2.4608.10. Wim Bogaerts acknowledges the Flemish Research Foundation (FWO-Vlaanderen) for a postdoctoral fellowship. 
%

%
\end{document}